# Genetics of single-cell protein abundance variation in large yeast populations

Frank W. Albert*, Sebastian Treusch, Arthur H. Shockley, Joshua S. Bloom and Leonid Kruglyak*

Howard Hughes Medical Institute
Lewis Sigler Institute for Integrative Genomics
Princeton University
Princeton, NJ 08544
USA

*corresponding authors:
FWA (falbert@princeton.edu) and LK (leonid@genomics.princeton.edu)

## Summary

Many DNA sequence variants influence phenotypes by altering gene expression. Our understanding of these variants is limited by sample sizes of current studies and by measurements of mRNA rather than protein abundance. We developed a powerful method for identifying genetic loci that influence protein expression in very large populations of the yeast *Saccharomyes cerevisiae*. The method measures single-cell protein abundance through the use of green-fluorescent-protein tags. We applied this method to 160 genes and detected many more loci per gene than previous studies. We also observed closer correspondence between loci that influence protein abundance and loci that influence mRNA abundance of a given gene. Most loci cluster at hotspot locations that influence multiple proteins—in some cases, more than half of those examined. The variants that underlie these hotspots have profound effects on the gene regulatory network and provide insights into genetic variation in cell physiology between yeast strains.

Variation among individuals arises in part from differences in DNA sequences, but the genetic basis for variation in most traits, including common diseases, remains only partly understood. A deeper understanding of the mechanisms by which variation in DNA sequence shapes phenotypes will aid progress in medical genetics, as well as provide insight into the workings of evolutionary change. Some DNA variants influence phenotypes by altering the expression level of one or multiple genes. The effects of such variants can be detected as expression quantitative trait loci (eQTL) [1]. eQTL can be divided into two classes based on their location in the genome [1]. Local eQTL lie close to the gene they influence and frequently act in *cis* in an allele-specific manner [2]. By contrast, distant eQTL can be located far from their target genes, and usually act in *trans* to affect both alleles of a gene. Studies utilizing designed crosses in model organisms have discovered an abundance of both local and distant eQTL [3-6]. Most studies in humans have focused on local eQTL in order to improve statistical power [7], although as sample sizes increase, distant loci are beginning to be unveiled [8-10].

Traditional eQTL mapping requires large-scale genotype and gene expression data for each individual in the study sample. Although generating such data sets is becoming easier and cheaper, this requirement has limited sample sizes to hundreds of individuals in both humans and model organisms [4-6,9-11]. Such studies are able to detect eQTL with strong effects, but are expected to miss many loci with smaller effects [12]. Further, the great majority of eQTL studies to date have used mRNA abundance as the measure of gene expression because of the relative ease and precision of microarray and sequence-based assays. However, coding genes exert their effects through their protein products. Following pioneering work using 2D gel electrophoresis [13,14], more recent studies have used mass-spectrometry proteomics to map QTL that influence protein levels (pQTL) [15-21]. These studies, which have been more limited in scope due to the difficulty of mass-spectrometry proteomics, found surprising differences between eQTL and pQTL for the same genes [16,17]. There is thus a clear need for new methods with simplified protein measurements as well as increased sample sizes and statistical power to study how genetic variation affects the proteome.

*GFP-based detection of loci that influence single-cell protein abundance*

We developed a method for detecting genetic influences on protein levels in large populations of genetically distinct individual yeast cells (Supplementary Figure S1). The method leverages extreme QTL mapping (X-QTL), a bulk segregant QTL mapping strategy with high statistical power [22]. We quantified protein abundance by measuring levels of green fluorescent protein (GFP) inserted in-frame downstream of a given gene of interest. The GFP tag allows protein abundance to be rapidly and accurately measured in millions of live, single cells by fluorescence-activated cell sorting (FACS). To apply the method to many genes, we took advantage of the yeast GFP collection [23,24], in which over 4,000 strains each contain a different gene tagged with GFP in a common genetic background (BY). For each gene under study, we crossed the GFP strain to a genetically divergent vineyard strain (RM) and generated a large pool of haploid GFP-positive offspring (segregants) of the same mating type. Across the genome, each segregant inherits either the BY or the RM allele at each locus, some of which influence the given gene's protein level. We used FACS to collect 10,000 cells each from the high and low tails of GFP levels (Supplementary Figure S2A). We then sequenced these extreme populations in bulk to a high depth of coverage that permitted accurate determination of allele frequencies, and detected loci that influence protein abundance as those genomic regions where the high and low GFP pools differ in the frequency of the parental alleles (Supplementary Figure S3). We denote these loci extreme protein QTL or X-pQTL.

*Protein levels are influenced by multiple loci*

We applied our method to 174 abundantly expressed genes, some of which were chosen based on previous eQTL and pQTL results, while others were selected at random. False discovery rates were determined using control experiments in which two cell populations were collected at random with respect to GFP level and compared as described above for high and low GFP pools (Supplementary Figure S2B). High-quality results were obtained for 160 genes (Supplementary Table S1 contains details on the genes, including exclusion criteria). Across these 160 genes, we identified 1,025 X-pQTL at a false discovery rate of 0.7% (Supplementary Dataset S1). The resulting X-pQTL were highly reproducible as gauged by biological and technical replicates (Supplementary Note 1 & Supplementary Figure S4). The number of X-pQTL per gene ranged from 0 to 25, with a median of five. We compared these numbers to eQTL and pQTL results based on published mRNA and mass-spectrometry protein data on ~100 segregants from a cross between the BY and RM strains [11,20]. 85 genes were assayed in all three data sets; for these genes, we observed a mean of 1.2 eQTL, 0.6 pQTL, and 7.2 X-pQTL (Figure 1A-B). Our method increased detection of loci that influence protein levels by an average of 1,100%, as compared to a 15% increase recently reported through improvements in mass spectrometry [25]. Interestingly, the distribution of detected loci per gene approaches that previously predicted to underlie gene expression variation in this yeast cross [12], indicating that the higher power of the method enabled discovery of loci with small effects that escaped detection in previous studies. Our detection of multiple X-pQTL per gene directly demonstrates considerable genetic complexity in protein expression variation.

*Abundant local variation in protein levels*

By design, all cells in the experiments described above inherit the GFP-tagged gene of interest, and the surrounding local region of the genome, from the BY strain. Therefore, the detected X-pQTL are distant from the gene of interest, and likely influence gene expression in *trans*. Previous studies have shown that mRNA levels of many genes are influenced by genetic variants in or near the gene itself (local eQTL), the majority of which influence gene expression in *cis* [1,2]. In contrast, local pQTL were reported to be rare based on mass spectrometry data [15,16]. We sought to investigate local X-pQTL by generating GFP-tagged strains for 41 genes in the RM genetic background, and repeating the crosses so that both parent strains carried GFP-tagged alleles, which then segregated among the progeny (Methods). We detected local X-pQTL at genome-wide significance for 20 genes (49%), and several additional genes would pass a more relaxed local significance threshold. Thus, local genetic variation affects protein levels for at least half of the genes we tested, in contrast to earlier observations [15,16] but in line with more recent work [18,21].

*Comparison of genetic influences on mRNA and protein levels*

The genetic architecture of protein and mRNA variation has previously been found to be surprisingly different [15-17]. For example, based on published microarray [11] and mass-spectrometry data [20], only 23% of the eQTL had a corresponding pQTL for the genes present in our current study (Figure 2B). By contrast, at least 53% of the distant eQTL have a corresponding X-pQTL for the genes shared between the mRNA [11] and our data set (Figure 2C). Further, the direction of QTL effect (i.e. whether higher expression is associated with the BY or the RM allele) agreed for 92% of these loci (randomization test $p < 0.01$). The local eQTL showed similar concurrence (13 of 21 local eQTL, or 62%, had a corresponding X-pQTL), with slightly lower directional agreement (77%). Thus, at least half of the loci with effects on mRNA levels also influence protein levels.

We next asked whether the many new X-pQTL discovered here (i.e., those without a corresponding significant eQTL) are best explained by posttranscriptional effects or by mRNA differences that are too small to be detected at genome-wide significance thresholds. In support of the latter hypothesis, we found that X-pQTL that correspond to significant eQTL have larger effects than those that do not (Figure 2D, Wilcoxon rank test, p = 6e-11). To further explore this question, we asked, at each X-pQTL position, if the published mRNA levels [11] of those individuals with the BY allele differed from those with the RM allele. From the distribution of the resulting p-values, we estimated [26] that at least 32% of the X-pQTL also influence mRNA levels. Although this result might suggest that 68% of X-pQTL arise from variants with posttranscriptional effects that do not influence mRNA levels, it is also consistent with variants that alter both mRNA and protein levels but explain < 1% of trait variance (Supplementary Note 2 and Supplementary Figure S5). Two recent studies which identified small sets of pQTL in humans [18] and yeast [21] also found substantial overlap with eQTL. Thus, the most parsimonious interpretation of our results is that the effects of most loci on protein levels arise from underlying mRNA variation.

Although genetic influences on mRNA and protein levels are overall more similar than previously suggested [16], we do observe instances of clear differences. For 21 of 109 significant distant eQTL (19%), we saw no evidence for a corresponding X-pQTL, even at low statistical stringency (LOD < 1). Five genes with strong local eQTL also showed no evidence for a local X-pQTL; conversely, four genes with a local X-pQTL had no corresponding eQTL (Table 1). These cases likely represent influences of genetic variation on posttranscriptional processes, although we cannot rule out subtle experimental differences between the mRNA and X-pQTL data sets.

*QTL hotspots with widespread overlapping effects on protein levels*

Distant eQTL in yeast, as well as in other species, are not randomly distributed throughout the genome, but instead cluster at "hotspot" loci that influence the expression of many genes [3,4,6]. We observed such clustering of X-pQTL at 20 genome locations, each of which overlapped more X-pQTL (≥ 12) than expected by chance (Table 2). The majority of all detected distant X-pQTL (69%) fell within a hotspot. Remarkably, these 20 X-pQTL hotspots captured nearly all of the mRNA hotspots identified in an eQTL data set for the same cross [11] (Figure 2A & Table 2). In contrast, many eQTL hotspots did not correspond to a mass-spectrometry based pQTL hotspot [16,20] (Figure 2B).

The X-pQTL hotspots had widespread effects on protein levels. The median fraction of genes a hotspot affected was 27% of the 160 genes tested, and two of the hotspots each affected more than half of the genes (Table 2). The magnitude and direction of expression changes differed considerably among the genes influenced by a given hotspot (Figure 3A). Together, these observations are best explained by hotspots shaping the proteome in a hierarchical manner. Proteins with the largest abundance changes are likely to be closely related in biological function to the gene whose alleles underlie a hotspot. Abundance of more distantly connected proteins may be more weakly perturbed through mechanisms that influence the overall physiological state of the cell [27].

The consequences of some genetic differences may thus reverberate through much of the cell. We illustrate these ideas with a closer look at three of the hotspots.

The hotspot at ~239 kb on chromosome XII influences the expression of nearly half the genes in our set (Figure 3B). It contains the gene *HAP1*, a transcriptional activator of genes involved in cellular respiration [28]. In BY, transcriptional activation by *HAP1* is reduced due to a transposon insertion, while *HAP1* function is intact in RM [3,29]. Of the nine genes in our dataset that are under direct transcriptional control by *HAP1* [30], seven were regulated by this hotspot (*YHB1*, *ACS2*, *CYC1*, *ERG10*, *OLE1*, *ADO1*, and *PDR16*), more than expected by chance (Fisher's Exact Test (FET) $p = 0.02$). Further, these seven direct *HAP1* targets all had reduced expression in the presence of the BY allele of *HAP1*, and they were more strongly influenced by the hotspot than the other genes linking here (Wilcoxon Test $p = 0.002$, Figure 3C). Similarly, the hotspot on chromosome XI contains the gene *HAP4*, which encodes a component of the Hap2/3/4/5 complex, an activator of respiratory gene expression with different target genes than *HAP1* [31]. Direct transcriptional targets of this complex [30] are enriched among the genes influenced by this hotspot in our data (5 / 6 genes, $p = 0.0003$), and these target genes were more strongly affected than other genes (Wilcoxon Test, $p = 0.02$). Notably, the BY allele was associated with lower expression at all these *HAP4* targets (Figure 3C). Thus, variation at both *HAP1* and *HAP4* regulates direct targets involved in cellular respiration. In both cases, the RM allele is associated with a more respiratory cellular state [27], likely resulting in the weaker expression changes for the many other genes affected by these hotspots.

The hotspot on chromosome XV regulates the largest fraction of genes in our dataset (Table 2). We previously showed that variation in the gene *IRA2* underlies the corresponding eQTL hotspot [11]. *IRA2* is an inhibitor of the Ras/PKA signaling pathway, which regulates a wide variety of processes, including the cellular response to glucose [32]. Addition of glucose to yeast growing on non-fermentable carbon sources results in expression changes at > 40% of all genes [32], and the majority of these changes are mediated through the Ras/PKA pathway [33]. The BY allele of *IRA2* is less active than the RM allele [11], and is therefore expected to be associated with higher Ras/PKA activity [27].

Indeed, the effects of this hotspot on protein levels are correlated with the mRNA expression changes induced by glucose addition [33] (Spearman rank correlation rho = 0.68, $p < 2e-16$, Figure 3D). The BY allele thus mimics stronger glucose signaling [27] even though glucose levels are constant and identical for all cells in our experiments. Interestingly, activation of respiratory genes by *HAP1* and *HAP4* is a branch of glucose signaling that is independent of Ras/PKA activity [33]. Thus, the BY laboratory strain differs from the wild RM strain in at least three key components of glucose sensing.

The hotspot effects often overlap for individual proteins. For example, the three hotspots described above jointly regulate a set of eleven genes in our dataset. The three BY alleles all reduced expression of five of these proteins (Table 3). Interestingly, these five genes (*ATP14*, *ATP17*, *ATP2*, *CIT1, MDH1*, see Supplementary Figure S6) are all involved in aerobic respiration, while the remaining six genes are not. The BY strain grows better than wild strains on glucose-rich media that favor fermentation over respiration [34,35]. Consistent direction of eQTL effects for genes in a pathway can be interpreted as evidence for adaptive evolution [36]. Thus, the *HAP1*, *HAP4* and *IRA2* hotspots may represent adaptations of BY to the glucose-rich culture conditions commonly used in the laboratory [37].

Ten X-pQTL hotspots did not have corresponding eQTL hotspots. They may arise from eQTL with effects below the detection limit of the earlier studies, or from variants that influence protein levels via posttranscriptional mechanisms. For example, the locus centered at 132,948 bp on chromosome II regulated about a third of genes in our dataset; the largest fraction among the 10 novel hotspots (Table 1). The BY allele increased expression of multiple ribosomal proteins and translation factors, suggesting that this hotspot regulates the abundance of ribosomes (Figure 3E & Supplementary Table S2). Interestingly, none of the ribosomal genes whose protein levels mapped to this hotspot had an eQTL at this locus, suggesting that it may influence ribosome abundance via posttranscriptional processes [38].

*Conclusions*

We developed a powerful method to detect genetic variants affecting protein levels and used it to uncover substantial complexity in gene expression regulation. We show that the genetic control of mRNA and protein levels is largely concordant. Individual proteins are typically influenced by multiple loci that cluster into hotspots with highly pleiotropic, overlapping effects. Our findings imply that many more eQTL and pQTL will be discovered in studies with larger sample sizes in other species, consistent with results that are beginning to emerge from human eQTL studies with many hundreds of individuals [8,9]. Our approach can be readily extended to any situation in which segregating cells can be subjected to fluorescent labeling and sorting.

# Materials and Methods

## Yeast Strains

We used strains from the yeast GFP collection [23] with genotype

*MATa his3Δ1 leu2Δ0 met15Δ0 ura3Δ0 GOI::GFP-HIS5*

where *GOI::GFP* signifies a carboxyterminal, in-frame insertion of the GFP gene to a gene of interest (GOI) [39]. All strains in the GFP collection have the same "BY" genetic background, a common laboratory strain. We crossed the GFP strains to one strain ("YLK2463") of the RM genetic background:

*MATα can1Δ::STE2pr-URA3-mCherry-KanMX his3Δ1::ClonNAT leu2Δ0 ura3Δ0 ho::HYG AMN1*$^{BY}$

YLK2463 carries the synthetic genetic array marker *STE2pr-URA3* [40] at the can1 locus that, in the presence of canavanine and the absence of uracil in the media, allows only cells of the '*a*' mating type to grow, permitting the rapid generation of large and stable segregant populations. The SGA marker was kindly provided by the laboratory of Charles Boone. We modified the SGA marker by adding a mCherry gene fused to the URA3 gene. Consequently, mCherry abundance is a measure of the expression of the SGA marker, permitting verification of successful selection of segregants. The BY strains and YLK2463 share the auxotrophies *his3Δ1*, *leu2Δ0* and *ura3Δ0* (but not *met15Δ0*) and carry identical alleles of the *AMN1* gene. Some of the strong *trans* eQTL identified in earlier mapping studies [3,11,12] were caused by engineered gene deletions (*leu2Δ0* and *ura3Δ0*) and by polymorphism at *AMN1*. Because these loci do not differ between our parent strains, the corresponding QTL do not occur in our experiments.

## Gene selection

We selected 174 genes for X-pQTL mapping from the ~4,000 genes represented in the GFP library (see Supplementary Table S1 for full information). Most genes (146) in our dataset were selected to have annotated GFP abundance > 300 in SD media [24]. Due to some genes being selected at random, 28 genes had published abundance lower than 300. 160 genes were represented in the Smith *et al.* eQTL dataset [11], and 102 genes were represented in the Khan *et al.* pQTL dataset [20]. Genes were further selected based on whether or not they had distant or local eQTL or pQTL. Among the 174 genes, 37 had a local eQTL, nine had a local pQTL, 101 had at least one distant eQTL and 30 genes had at least one distant pQTL.

In this paper, we present data from 160 of these 174 genes. The remaining genes were excluded due to poor growth of the GFP-tagged strain leading to either no useable data, or to insufficient sequencing data. For five genes, we replaced failed "trans" experiments with those from the "local" experiments (s. below). Two of these five genes had local X-pQTL. The inclusion of these two local loci in the 1,025 X-pQTL discussed in the paper does not alter our conclusions. All details on gene selection and exclusion criteria are given in Supplementary Table S1.

## Generation of pools of segregant offspring

For each cross, YLK2463 and the corresponding BY strain from the GFP collection were mated and diploids selected on YNB + Leu + Ura + Hygromycin plates. Diploid cultures were sporulated for ~7 days in 5ml Spo++ media. Spores were plated on YNB + Leu + Met + Canavanine plates. The presence of canavanine and the absence of uracil select for both the deletion of *CAN1* by the SGA marker and for cells of the '*a*' mating type (i.e. the BY allele in our cross). The absence of histidine selected for the presence of the GFP cassette, ensuring that all surviving segregants carry the fluorescently labeled allele of the gene of interest. Segregants were harvested after two days, and glycerol stocks frozen at -80°C. Successful selection of MAT *a* cells that carry both GFP and the active magic marker was verified during FACS sorting by the presence of both GFP and mCherry signal.

For the local pQTL experiments, both parent strains are histidine prototroph and therefore diploids cannot be selected for chemically. Instead, diploids were manually picked from freshly mated cultures using a yeast tetrad dissection scope (MSM System from Singer Instruments, Somerset, UK).

**Fluorescence-activated cell sorting**

Segregant libraries were thawed and grown for ~12 h in 5 ml of selective media (YNB + Leu + Met + Canavanine). Cells were directly FACS sorted from and into culture media, with no intermediate exposure to nutrient-free buffers. FACS sorts were performed on a BD FACSVantage SE instrument (BD Biosciences, Franklin Lakes, NJ, USA). For each experiment, 10,000 cells were collected from the populations with the 1-2% highest and lowest GFP signal respectively, while controlling for cell size as measured by forward scatter (Supplementary Figure S2). All isolated populations were grown for ~30 h in liquid YNB Leu + Met + Canavanine media and frozen at -80 °C as glycerol stocks. For all downstream procedures, the high and low populations were treated identically, and processed at the same time.

**Empirical estimates of the false discovery rate**

In QTL mapping studies involving individual segregants, the false discovery rate (FDR) is typically determined by permuting phenotypes relative to genotype data. This is not possible in X-QTL as in this approach, the genotypes of individual segregants are not known. Instead, we determined the distribution of random allele frequency fluctuations that can occur without selection on GFP levels. We grew two replicates each of segregant pools for 10 genes and one replicate for one additional gene, for a total of 21 experiments. In each experiment, we selected two populations of 10,000 cells in the same cell size range as for the GFP sorts, but without gating on GFP abundance (Supplementary Figure S2). The resulting 21 pairs of 10,000 cells were then processed and sequenced exactly as described for high / low GFP populations. We applied our peak calling pipeline (s. below) to the data from these 21 experiments and determined the number of loci that would be called significant at a range of thresholds. We set the genome-wide threshold of LOD = 4.5 for further analyses to the highest LOD score (when incrementing in steps of 0.1 LOD) where we see one QTL across the 21 "null" experiments.

**DNA library preparation and sequencing**

High and low pools were thawed and about 30% grown for ~12 h in YNB + Leu + Met + Canavanine. DNA was extracted using the Quiagen DNEasy system. Indexed, paired-end Illumina libraries were constructed from 25 ng of genomic DNA, using a modification of the Epicentre Nextera [41] protocol using 20X diluted tagmentation enzyme [42] and 11 cycles of post-tagmentation PCR. We used a set of 96 custom Nextera-compatible adaptor primers that contain index sequences described in [43]. Up to 96 indexed samples (corresponding to 48 pairs of high and low GFP pools) were pooled to equal molarity and size selected on agarose gels to 400 – 500bp length. Sequencing was performed on an Illumina HiSeq 2000 instrument (Illumina Inc, San Diego, CA, USA), using a read length of 100 bp, with some library pools sequenced as single end and others as paired end. Sequencing depth ranged from 15X – 68X coverage of the whole genome, with a median of 34X. Raw sequencing reads are available upon request.

**Measuring allele frequencies by massively parallel short-read sequencing**

BY and RM differ at ~0.5% of nucleotides, corresponding to ~ 45,000 single nucleotide variants (SNPs) that can serve as dense genetic markers in QTL mapping experiments [22,42]. A challenge for accurate estimation of allele frequencies is mapping bias, i.e. a systematic tendency for sequencing reads corresponding to the reference strain to map better than reads that contain alleles from a non-reference strain. Mapping bias is of particular concern in our experiments because the yeast reference genome was generated from one of our strains (BY). We initially noted clear evidence of reference bias in our data, even though our reads were comparably long. We therefore took several steps to eliminate mapping bias.

First, we compiled a catalogue of high-quality SNPs from Illumina genomic sequence data of the BY and the RM strain [42]. Second, we restricted this catalogue to SNPs that can be unambiguously aligned to RM by making use of the high quality RM reference genome that is available from the Broad Institute (http://www.broadinstitute.org/annotation/genome/saccharomyces_cerevisiae.3/Info.html ). For each SNP, we extracted 30 bp up- and downstream sequence from the BY reference, and set the SNP position itself to the RM allele. We aligned the resulting 61 bp fragments (as well as their reverse complement) to the RM genome using BWA [44]. We kept only SNPs where both the forward and the reverse "read" aligned uniquely to RM, resulting in a set of 38,430 SNPs. Third, we aligned the reads from each experiment to both the BY and the RM reference using BWA [44]. At each SNP, we kept only reads that mapped uniquely and without mismatches. Thus, reads that span a SNP were only retained when mapped to the strain reference from which they originated. While we acknowledge that this procedure removes reads with sequencing errors, we found that the corresponding loss in sequence coverage was justified by the improved accuracy of allele frequency estimates. Finally, we removed likely PCR duplicates using a python script kindly provided by Martin Kircher, and estimated allele frequencies by counting at each SNP the number of reads that matched the BY or the RM references. Together, these procedures resulted in dense, accurate allele frequency estimates across the entire yeast genome.

Allele count data used in downstream analyses will be made available upon publication.

**Analyses of count data and QTL detection**

Unless otherwise specified, all statistical analyses were performed in the R programming environment (www.r-project.org). At each SNP, we calculated two statistics to describe the allele frequency distribution in the pools. First, we simply calculated the fraction of reads with the BY allele in each pool and subtracted these frequencies in the low GFP tail from those in the high GFP tail ("allele frequency difference"). Second, we calculated the p-value from a G-test comparing the number of BY vs. RM counts in the high to those in the low GFP tail. Because these two SNP-wise statistics can be highly variable at neighboring SNPs due to random sampling, we performed loess-smoothing along the chromosomes for plotting results for single genes.

For X-pQTL detection, we used the MULTIPOOL software [45]. MULTIPOOL fits a graphical model to each chromosome that takes into account both linkage and variation in sequence coverage. MULTIPOOL reports a LOD score from a likelihood ratio test comparing a model with and a model without a QTL at the given position. MULTIPOOL was run in "contrast" mode, and with the following parameters: base pairs per centiMorgan (-c parameter) = 2200, bin size (-r) = 100. The pool size (-n) was set to 1,000 rather than 10,000 to allow for the fact that not all collected cells will survive. We noticed that MULTIPOOL can be highly sensitive to SNPs that are fixed or nearly fixed for one of the parental alleles. At these positions, MULTIPOOL sometimes produces very sharp peaks in the LOD curve that spike at single SNPs. We therefore removed SNPs with a BY allele frequency $> 0.9$ or $< 0.1$ prior to running MULTIPOOL. The resulting LOD curves robustly detect peaks, and are free from any single-SNP artifacts.

We used the empirical null sorts to set the genome-wide threshold for peak detection at a LOD $\geq 4.5$ (s above). Within each QTL, we considered the position of the highest LOD score, and defined confidence intervals as the 2-LOD drop interval around this peak. For a given LOD threshold, false discovery rates were estimated as

[# QTL in the 21 null sorts * (# experiments/21)] / # QTL

Finally, as a measure of the effect size of an X-pQTL, we used the loess-smoothed allele frequency difference between the high and the low GFP population.

**Measuring library purity from sequence data**

To ensure that each of our experiments targeted the intended gene of interest, and were free from cross-experiment or cross-library contamination, we made use of the fact that deep sequence data allows direct detection of the gene in an experiment that are tagged by the GFP cassette. We created a reference fasta file with two sequences for each gene as follows. First, we added the terminal 75 bp of the gene's ORF sequence immediately upstream of (but excluding) the stop codon to the first 75 bp of the GFP cassette. Second, we added the last 75 bp of the cassette to the 75 bp of genomic sequence immediately downstream of the stop codon. The cassette sequence was obtained from http://yeastgfp.yeastgenome.org. The length of the sequences were chosen such that a 100 bp read can only map to them if it contains the point of insertion of the GFP cassette.

We mapped all reads to this fusion reference, treating paired reads as single reads (because if two paired reads are mapped as such, only one of them can perfectly cover the insertion site, while the second read is not informative in this context). We used the samtools idx tool to count the number of reads that mapped to each fusion sequence, allowing direct identification of the tagged gene and quantification of any off-target reads.

With a few exceptions (discussed below), all experiments reported here were > 90% pure for the gene of interest in both the high and the low GFP pool. Off-target reads typically corresponded to other genes in the study, suggesting that they may be due to either low levels of cross-contamination during library preparation in 96 well format, or incorrectly sequenced indeces.

We noticed two clear outliers in terms of estimated purity. First, the pools for gene YDR343C (*HXT6*) had 50-60% of reads mapping to the gene YDR342C (*HXT7*). These two genes are close paralogs, and both the ends of their ORFs and their downstream sequence are virtually identical, suggesting that the apparent contamination is in fact due to reads randomly mapping to either of the two genes. Second, the pools for the gene YGR192C (*TDH3*) appeared to be ~20 – 25% contaminated by the gene

YGR009C (*TDH2*). These two genes are also close paralogues so that the ends of their ORFs used in our fusion library are identical, but have different downstream sequences. Off-target reads are therefore expected at 25% of reads for YGR192C. We retained both YDR343C and YGR192C in our analyses.

The remaining genes with apparent contamination have low absolute numbers of reads overlapping the cassette fusions so that a single off-target read has a disproportionate effect on the purity estimate. The one exception is YBR158W (*AMN1*), where 7 out of 68 fusion reads in the high GFP tail mapped to the gene YIL043C (*CBR1*), which is not a paralog of YBR158W. We removed YBR158 from all further analyses.

**Detecting local X-pQTL**

In the experiments described so far, all segregants carry the GFP cassette only at the BY allele of the gene of interest, so that we can detect only distant X-QTL. To test the effect of local variation on a given gene, we engineered the corresponding GFP cassette into our RM strain YLK2463. The GFP cassette along with the HIS5 gene was amplified from genomic DNA extracted from the respective GFP collection strain using primers designed using sequences available at http://yeastgfp.yeastgenome.org/yeastGFPOligoSequence.txt. YLK2463 was transformed with the PCR product and transformants selected on HIS- media following standard yeast protocols. Successful integration at the carboxyterminal end of the target gene was verified using colony PCR with primers described in [39]. Because the alleles from both parents are now tagged with GFP, these experiments allow the detection of local X-QTL. We mapped X-pQTL as described above.

We selected 55 genes to be included in the "local" experiments based on whether or not they had a local eQTL or pQTL [11,20] and whether or not they showed allele-specific expression in RNASeq experiments (Albert *et al*. unpublished, Torabi *et al*. unpublished). All 55 genes were also included in the 174 "distant" experiments described above (Supplementary Table S1). We excluded seven "local" experiments due to low growth or insufficient sequencing data. The "distant" experiments were FACS sorted and further processed at the same time as the "local" experiments, allowing direct comparison of the results.

To ensure that the GFP cassette is intact after transformation, we analyzed alignments from the high and low GFP populations against the GFP gene sequence. We detected several GFP mutations that were in common between the RM strain and the donor GFP from the corresponding BY strain and that were therefore already present in the GFP collection strains. At five genes, the RM strain carried silent mutations that were not found in the BY strain; these are unlikely to cause false positive local X-pQTL and the genes were retained in our analyses. We excluded six genes with nonsynonymous mutations present only in the RM allele where the RM allele was associated with lower GFP fluorescence. For three genes, we noted non-synonymous mutations in the RM GFP sequence where the RM allele associated with higher fluorescence. Because a fortuitous mutation in the GFP ORF is unlikely to increase GFP fluorescence, it is unlikely that the mutations alter the GFP signal in these three cases. These three genes were therefore retained in the analyses. Two of these three genes (YKL029C and YNL061W) had a local X-pQTL with concordant expression direction to a local eQTL, while the third gene (YBR067C) had a local X-pQTL and no data available in the eQTL dataset. Finally, we excluded one gene where the GFP cassette had no mutations, but where several sequencing reads spanned the end of the ORF without being interrupted by the GFP cassette, suggesting that not all segregants may have inherited a GFP-tagged allele. Supplementary Table S1 details all gene exclusions.

In the paper, we present data for the 41 genes with high quality data. Matched *trans* data was available for 37 of these genes. A local X-pQTL is called if the LOD score at the midpoint of the gene exceeds a given threshold (e.g. LOD > 4.5 for genome-wide significance). Because genome-wide significance is conservative when assaying only a single position in the genome as for local X-pQTL, we also used a more relaxed local significance threshold. This threshold was set to the maximum LOD score at the gene position in the "null" experiments described above (LOD = 0.8).

## eQTL mapping from published datasets

We obtained genome-wide microarray based gene expression measures from Smith & Kruglyak [11], as well as mass-spectrometry based protein quantifications from Khan et al. [20]. Because these data were measured in the same set of ~100 segregants (albeit at different points in time and therefore from separate cultures), we can analyze them in an identical fashion using the available set of genotypes for these segregants [11]. We performed nonparametric linkage mapping using R/QTL [46] for each gene, and called QTL at a threshold of LOD = 3, with confidence intervals defined as the 2-LOD drop from the peak position. We note that this is not a stringent cutoff in an eQTL experiment where multiple traits are mapped. However, because we compare these peaks to those from our X-QTL approach (which are controlled for multiple testing using an FDR approach), being more permissive here in fact downplays the improvements in QTL detection by our method.

## Clustering of X-QTL into hotspots

To determine if the X-pQTL were non-randomly distributed across the genome, we reshuffled the observed X-pQTL peak positions 100 times across the genome. In the randomizations, each chromosome was sampled with a probability proportional to its length, and the sizes of confidence intervals were kept intact. In each set of randomized loci, we counted for each SNP the number of overlapping X-pQTL (using 2-LOD confidence intervals). The cutoff for "significant" hotspots was set to the median of the 95% quantiles from the 100 randomized sets.

To identify individual hotspots, we extracted continuous stretches of SNPs that match or exceed the empirical cutoff determined above. Stretches of less than six SNPs were excluded. Within each of the remaining stretches, we defined the hotspot position to be the SNP that overlapped the most X-pQTL (defined by 2-LOD drop confidence intervals). If multiple SNPs overlapped the same number of X-pQTL, we selected the SNP with the smallest bp position to be the hotspot position.

Note that Figure 2 groups linkages into bins of 20 cM (based on the linkage map used in [42]). The threshold displayed in that figure is based on 100 randomizations of peak positions as described above, but was not itself used for determining hotspot locations. We chose this visual display to be consistent with that in earlier work [11]. The hotspots that identified using the method described above are identical to those that would be identified using 20 cM bins, as can be seen by comparing Table 2 with Figure 2.

**Overlap of eQTL with X-pQTL or pQTL**

For each eQTL, we asked if it was located within 44 kb (roughly 20 cM) of an X-pQTL or pQTL for the given gene. In the published eQTL and pQTL datasets, we defined peaks as those loci exceeding a LOD threshold of ≥ 3. We excluded loci that are known to segregate in only one of the datasets: in particular, we removed the following eQTL from the published dataset before comparing to the X-pQTL data:

- All eQTL on chromosome II (due to polymorphism in the gene *AMN1* [47]; our RM strain carries the BY allele of *AMN1* so that this locus cannot influence protein expression in our data)
- All eQTL on chromosome III (due to the mating type locus [3] which is identical in all our segregants, or to an engineered auxotrophy in the gene *LEU2* [3] which was present only in BY in the earlier data, while *LEU2* is deleted in both of our parent strains)
- All eQTL on the chromosome where the gene itself is located because in our "distant" experiments, all segregants share the GFP-tagged BY allele and hence local effects cannot be detected

We note that this strategy will remove a small set of loci that are located on excluded chromosomes but do not correspond to the loci specified above. Excluding these loci is unlikely to influence our overall conclusions. When comparing eQTL with mass-spectrometry based pQTL, we retained all loci in the analyses because the segregants used in these two studies are for the most part identical, so that the same loci are expected to be present in both datasets. Further, when comparing eQTL and pQTL we only analyzed genes that are included in the X-pQTL dataset, to avoid any biases related to the gene selection. If all genes shared between Smith *et al.* [11] and Khan *et al.* [20] are analyzed, there are 504 eQTL, only 62 of which are also pQTL (12%). Therefore, restricting the overlap analyses to genes present in the X-pQTL dataset leads to a better agreement between the earlier eQTL and pQTL datasets than across all genes, and doing so is conservative for our purposes.

We further asked if the direction of effect for an X-pQTL agrees with that for an overlapping eQTL. For example, at a given locus, a higher frequency of the BY allele in the high GFP tail compared to the low GFP tail was interpreted as the BY allele increasing protein expression. This measure was compared to the difference in measured mRNA expression between those segregants that inherited the BY vs. those that inherited the RM allele among the ~100 segregants in the published datasets.

To test if the observed overlap and directional agreement between X-pQTL and mRNA eQTL exceeded that expected by chance, we performed a randomization test. While leaving the positions of X-pQTL and their associated allele frequencies intact for a given gene (i.e. without redistributing X-pQTL across the genome, and without redistributing them between genes), we randomly re-assigned gene names to the gene-wise sets of X-pQTL positions and allele frequencies. From each of 100 randomized sets, we calculated the number of times an mRNA eQTL overlaps a directional effect with an X-pQTL, and what fraction of the overlapping QTL have an effect in the same direction. This test is conservative because of the presence of the linkage hotspots: because many genes link here in both the X-pQTL and eQTL data, a high degree of random overlap is expected. Our test asked whether the observed degree of gene-by-gene overlap exceeds even this high background expectation.

The result of the procedure described so far is the fraction of eQTL that overlap an X-pQTL or a pQTL. Because X-pQTL are so abundant, a potential concern is that this fraction (while higher than expected by chance, see main text) could be inflated due to chance overlap with non-specific X-pQTL. To guard against this possibility, we performed 100 randomizations of eQTL positions as described above for the clustering analyses. In each randomized dataset, we extracted the eQTL / X-pQTL overlap fraction.

The main text reports the observed fraction (58%) less the mean of these 100 randomized fractions (5%). An analogous correction was applied to the overlap of eQTL with mass-spectrometry pQTL (observed fraction = 23.5%; randomized mean = 0.5%).

## Testing the effect of X-pQTL on mRNA levels

In these analyses, we used the mRNA expression data across ~100 BY / RM segregants reported by Smith & Kruglyak [11]. We restricted the analyses to the 793 X-pQTL that are not located on chromosomes II and III, and also excluded, for each gene, the chromosome on which the gene is located. For each X-pQTL, we obtained the mRNA levels of the given gene in those segregants with the BY and those with the RM allele. We then performed a T-test comparing these mRNA levels and recorded the p-values. The p-value distribution was used to compute $\pi_0$, the fraction of true negative tests and $\pi_1 = 1-\pi_0$, a lower bound for the fraction of true positive tests [26]. $\pi_1$ provides a lower bound for the fraction of X-pQTL that affect mRNA levels. We used the R package qvalue [26] for these calculations.

Because of the large number of X-pQTL, we sought to correct the $\pi_1$ estimate for the expectation if random loci in the genome are sampled. We randomized the X-pQTL as described above for 100 times, each time calculating $\pi_1$. The mean $\pi_1$ across the randomized datasets (9.8%) was then subtracted from the estimate from the real data (42%) to arrive at the figure provided in the Results.

## Comparison of genes regulated by hotspots to other datasets

### *HAP1 and HAP4 targets*

Genes regulated by the *HAP1* and *HAP4* transcription factors were downloaded from ScerTF [48], using ChIP data for both transcription factors. Overlap between transcription factor targets and the genes regulated by the given hotspots was tested using Fisher's exact test. Effect sizes for a gene at a hotspot position were measured as the difference in allele frequency of the BY allele between the high and low GFP population. Effect sizes for transcription factor targets and the remaining genes were tested using Wilcoxon rank tests.

*Expression data for glucose sensing and PKA induction*

To test if the putative *IRA2* hotspot mimics the effects of altered glucose sensing, we compared the effects of this locus on the genes in our dataset to mRNA expression data obtained by Zaman et al. [33]. In that work, the authors added glucose to yeast growing on glycerol (a non-fermentable carbon source) and measured the resulting mRNA expression changes using microarrays. We obtained these expression data from the PUMA database (http://puma.princeton.edu). We averaged the results for each gene across the four available replicates of the 60 minutes time point post glucose addition (experiment IDs 100564, 101022, 101261, 105490). We calculated spearman's rank correlation between hotspot effect size and mRNA expression. The hotspot effects are polarized such that positive values correspond to higher expression being caused by the BY compared to the RM allele.

## References

1. Rockman, M. V. & Kruglyak, L. Genetics of global gene expression. *Nature Reviews Genetics* **7,** 862–872 (2006).
2. Ronald, J., Brem, R. B., Whittle, J. & Kruglyak, L. Local Regulatory Variation in Saccharomyces cerevisiae. *PLoS Genetics* **1,** e25 (2005).
3. Brem, R. B., Yvert, G., Clinton, R. & Kruglyak, L. Genetic Dissection of Transcriptional Regulation in Budding Yeast. *Science* **296,** 752–755 (2002).
4. Rockman, M. V., Skrovanek, S. S. & Kruglyak, L. Selection at linked sites shapes heritable phenotypic variation in C. elegans. *Science* (2010).
5. Huang, G. J. *et al.* High resolution mapping of expression QTLs in heterogeneous stock mice in multiple tissues. *Genome Research* **19,** 1133–1140 (2009).
6. West, M. A. L. *et al.* Global eQTL Mapping Reveals the Complex Genetic Architecture of Transcript-Level Variation in Arabidopsis. *Genetics* **175,** 1441–1450 (2006).
7. Gaffney, D. J. Global Properties and Functional Complexity of Human Gene Regulatory Variation. *PLoS Genetics* **9,** e1003501 (2013).
8. Rotival, M. *et al.* Integrating Genome-Wide Genetic Variations and Monocyte Expression Data Reveals Trans-Regulated Gene Modules in Humans. *PLoS Genetics* **7,** e1002367 (2011).
9. Grundberg, E. *et al.* Mapping cis- and trans-regulatory effects across multiple tissues in twins. *Nature Genetics* **44,** 1084–1089 (2012).
10. Emilsson, V. *et al.* Genetics of gene expression and its effect on disease. *Nature* **452,** 423–428 (2008).
11. Smith, E. N. & Kruglyak, L. Gene–Environment Interaction in Yeast Gene Expression. *PLoS Biology* **6,** e83 (2008).
12. Brem, R. B. & Kruglyak, L. The landscape of genetic complexity across 5,700 gene expression traits in yeast. *Proceedings of the National Academy of Sciences* **102,** 1572–1577 (2005).
13. Klose, J. *et al.* Genetic analysis of the mouse brain proteome. *Nature Genetics* **30,** 385–393 (2002).
14. Damerval, C., Maurice, A., Josse, J. M. & de Vienne, D. Quantitative trait loci underlying gene product variation: a novel perspective for analyzing regulation of genome expression. *Genetics* **137,** 289–301 (1994).
15. Foss, E. J. *et al.* Genetic basis of proteome variation in yeast. *Nature Genetics* **39,** 1369–1375 (2007).
16. Foss, E. J. *et al.* Genetic Variation Shapes Protein Networks Mainly through Non-transcriptional Mechanisms. *PLoS Biology* **9,** e1001144 (2011).
17. Ghazalpour, A. *et al.* Comparative Analysis of Proteome and Transcriptome Variation in Mouse. *PLoS Genetics* **7,** e1001393 (2011).
18. Wu, L. *et al.* Variation and genetic control of protein abundance in humans. *Nature* (2013). doi:10.1038/nature12223
19. Holdt, L. M. *et al.* Quantitative Trait Loci Mapping of the Mouse Plasma Proteome (pQTL). *Genetics* **193,** 601–608 (2012).

20. Khan, Z., Bloom, J. S., Garcia, B. A., Singh, M. & Kruglyak, L. Protein quantification across hundreds of experimental conditions. *Proceedings of the National Academy of Sciences* **106,** 15544–15548 (2009).
21. Skelly, D. A. *et al.* Integrative phenomics reveals insight into the structure of phenotypic diversity in budding yeast. *Genome Research* (2013). doi:10.1101/gr.155762.113
22. Ehrenreich, I. M. *et al.* Dissection of genetically complex traits with extremely large pools of yeast segregants. *Nature* **464,** 1039–1042 (2010).
23. Huh, W.-K. *et al.* Global analysis of protein localization in budding yeast. *Nature* **425,** 686–691 (2003).
24. Newman, J. R. S. *et al.* Single-cell proteomic analysis of S. cerevisiae reveals the architecture of biological noise. *Nature* **441,** 840–846 (2006).
25. Picotti, P. *et al.* A complete mass-spectrometric map of the yeast proteome applied to quantitative trait analysis. *Nature* **494,** 266–270 (2013).
26. Storey, J. D. & Tibshirani, R. Statistical significance for genomewide studies. *Proceedings of the National Academy of Sciences* **100,** 9440–9445 (2003).
27. Litvin, O., Causton, H. C., Chen, B. J. & Pe'er, D. Modularity and interactions in the genetics of gene expression. *Proceedings of the National Academy of Sciences* **106,** 6441–6446 (2009).
28. Zitomer, R. S. & Lowry, C. V. Regulation of gene expression by oxygen in Saccharomyces cerevisiae. *Microbial Reviews* **56,** 1–11 (1992).
29. Gaisne, M., Bécam, A. M., Verdiere, J. & Herbert, C. J. A ‘natural’ mutation in Saccharomyces cerevisiae strains derived from S288c affects the complex regulatory gene HAP1 ( CYP1 ). *Current Genetics* **36,** 195–200 (1999).
30. Harbison, C. T. *et al.* Transcriptional regulatory code of a eukaryotic genome. *Nature* **431,** 99–104 (2004).
31. Butler, G. Hypoxia and Gene Expression in Eukaryotic Microbes. *Annual Review of Microbiology* **67,** 291–312 (2013).
32. Zaman, S., Lippman, S. I., Zhao, X. & Broach, J. R. How Saccharomyces Responds to Nutrients. *Annu. Rev. Genet.* **42,** 27–81 (2008).
33. Zaman, S., Lippman, S. I., Schneper, L., Slonim, N. & Broach, J. R. Glucose regulates transcription in yeast through a network of signaling pathways. *Mol Syst Biol* **5,** – (2009).
34. Spor, A. *et al.* Niche-driven evolution of metabolic and life-history strategies in natural and domesticated populations of Saccharomyces cerevisiae. *BMC Evol Biol* **9,** 296 (2009).
35. Warringer, J. *et al.* Trait Variation in Yeast Is Defined by Population History. *PLoS Genetics* **7,** e1002111 (2011).
36. Fraser, H. B., Moses, A. M. & Schadt, E. E. Evidence for widespread adaptive evolution of gene expression in budding yeast. *Proceedings of the National Academy of Sciences* **107,** 2977–2982 (2010).
37. Lewis, J. A. & Gasch, A. P. Natural Variation in the Yeast Glucose-Signaling Network Reveals a New Role for the Mig3p Transcription Factor. *G3 - Genes|Genomes|Genetics* **2,** 1607–1612 (2012).
38. Henras, A. K. *et al.* The post-transcriptional steps of eukaryotic ribosome biogenesis. *Cell. Mol. Life Sci.* **65,** 2334–2359 (2008).

39. Howson, R. *et al.* Construction, Verification and Experimental Use of Two Epitope-Tagged Collections of Budding Yeast Strains. *Comparative and Functional Genomics* **6,** 2–16 (2005).
40. Tong, A. H. Y. & Boone, C. High-Throughput Strain Construction and Systematic Synthetic Lethal Screening in *Saccharomyces cerevisiae*. *Methods in Microbiology* **36,** 369–707 (2007).
41. Adey, A. *et al.* Rapid, low-input, low-bias construction of shotgun fragment libraries by high-density in vitro transposition. *Genome Biol* **11,** R119 (2010).
42. Bloom, J. S., Ehrenreich, I. M., Loo, W. T., Lite, T.-L. V. & Kruglyak, L. Finding the sources of missing heritability in a yeast cross. *Nature* **494,** 234–237 (2013).
43. Meyer, M. & Kircher, M. Illumina Sequencing Library Preparation for Highly Multiplexed Target Capture and Sequencing. *Cold Spring Harbor Protocols* (2010). doi:10.1101/pdb.prot5448
44. Li, H. & Durbin, R. Fast and accurate short read alignment with Burrows-Wheeler transform. *Bioinformatics* **25,** 1754–1760 (2009).
45. Edwards, M. D. & Gifford, D. K. High-resolution genetic mapping with pooled sequencing. *BMC Bioinformatics* **13,** S8 (2012).
46. Broman, K. W., Wu, H., Sen, S. & Churchill, G. A. R/qtl: QTL mapping in experimental crosses. *Bioinformatics* **19,** 889–890 (2003).
47. Yvert, G. *et al.* Trans-acting regulatory variation in Saccharomyces cerevisiae and the role of transcription factors. *Nature Genetics* **35,** 57–64 (2003).
48. Spivak, A. T. & Stormo, G. D. ScerTF: a comprehensive database of benchmarked position weight matrices for Saccharomyces species. *Nucleic Acids Research* **40,** D162–D168 (2011).

## Acknowledgements

We are grateful to Christina deCoste in the Princeton Flow Cytometry Resource Facility for technical assistance and advice on the experiments. This work was supported by National Institutes of Health (NIH) grant R01 GM102308, a James S. McDonnell Centennial Fellowship, and the Howard Hughes Medical Institute (LK), a research fellowship from the German Science Foundation AL 1525/1-1 (FWA), a National Science Foundation (NSF) fellowship (JSB), and NIH postdoctoral fellowship F32 GM101857-02 (ST)

## Author contributions

FWA and LK conceived the project, designed research and wrote the paper. FWA and AHS performed experiments. FWA analyzed the data. ST provided advice on yeast strain construction, the initial experimental design and other experimental procedures. JSB provided advice on experimental procedures and data analysis.

## Competing financial interest statement

The authors declare that no competing financial interests exist.

Correspondence and requests for materials should be addressed to FWA (falbert@princeton.edu) and LK (leonid@genomics.princeton.edu).

## Tables

Table 1 – mRNA-specific and protein-specific local QTL

| Gene | X-pQTL LOD | eQTL LOD |
|---|---|---|
| Local eQTL only | | |
| YJL201W | 0.5 | 15.2 |
| YPL048W | 0.4 | 7.3 |
| YDL171C | 0.5 | 6.4 |
| YLR438W | 1.0 | 6.4 |
| YNL044W | 0.5 | 5.3 |
| Local X-pQTL only | | |
| YJL130C | 6.4 | 0.2 |
| YDL126C | 13.7 | 0.2 |
| YGL026C | 8.6 | 0.1 |
| YMR315W | 12.7 | 0.6 |

Table 2 – Hotspot regulators of protein expression

| chromosome | Position (peak SNP) | % of genes regulated at LOD > 4.5 / LOD > 3 | mRNA hotspot[1] |
|---|---|---|---|
| I | 39,010 | 31 / 40 | Glu1 |
| II | 132,948 | 31 / 41 | - |
| II | 397,978 | 9 / 18 | Glu2 |
| IV | 223,943 | 12 / 24 | - |
| V | 192,064 | 16 / 31 | - |
| V | 371,845 | 16 / 21 | Glu6 |
| VII | 137,332 | 15 / 26 | - |
| VII | 505,871 | 16 / 29 | - |
| VIII | 103,041 | 19 / 29 | Glu7 |
| VIII | 419,747 | 8 / 12 | - |
| X | 142,009 | 18 / 26 | - |
| X | 655,465 | 11 / 15 | - |
| XI | 234,462 | 16 / 23 | Glu8 |
| XII | 238,302 | 16 / 31 | - |
| XII | 656,893 | 41 / 49 | Glu9 |
| XII | 1,039,502 | 12 / 19 | Yvert[2] |
| XIII | 96,832 | 31 / 46 | Glu10 |
| XIV | 232,509 | 13 / 19 | - |
| XIV | 465,007 | 58 / 65 | Glu11 |
| XV | 162,766 | 56 / 70 | Glu12 |

[1]As identified in Smith & Kruglyak 2008 [11].

[2]This hotspot was not observed in Smith & Kruglyak [11], but was present in an earlier BY / RM eQTL dataset[47].

Table 3 – Genes regulated by the four hotspots discussed in the text

| Gene | chrXI effect | chrXII effect | Chr XV effect | Description |
|---|---|---|---|---|
| *ATP14** | -0.35 | -0.14 | -0.14 | ATP synthase |
| *ATP17** | -0.14 | -0.14 | -0.18 | ATP synthase |
| *ATP2** | -0.21 | -0.3 | -0.22 | ATP synthase |
| *CIT1** | -0.23 | -0.36 | -0.26 | Citrate synthase |
| *MDH1** | -0.22 | -0.1 | -0.39 | Malate Dehydrogenase |
| *ADO1* | -0.09 | -0.25 | 0.09 | Adenosine kinase |
| *GLT1* | -0.08 | 0.13 | 0.24 | Glutamate synthase |
| *LIA1* | -0.1 | 0.15 | 0.15 | Deoxyhypusine hydroxylase |
| *TDH3* | -0.14 | 0.35 | 0.27 | Glyceraldehyde-3-phosphate dehydrogenase (GAPDH) |
| *YHB1* | -0.16 | -0.92 | 0.13 | Nitric oxide oxidoreductase |
| *YLR179C* | -0.09 | 0.7 | 0.17 | Unknown function |

* involved in aerobic respiration

# Figures

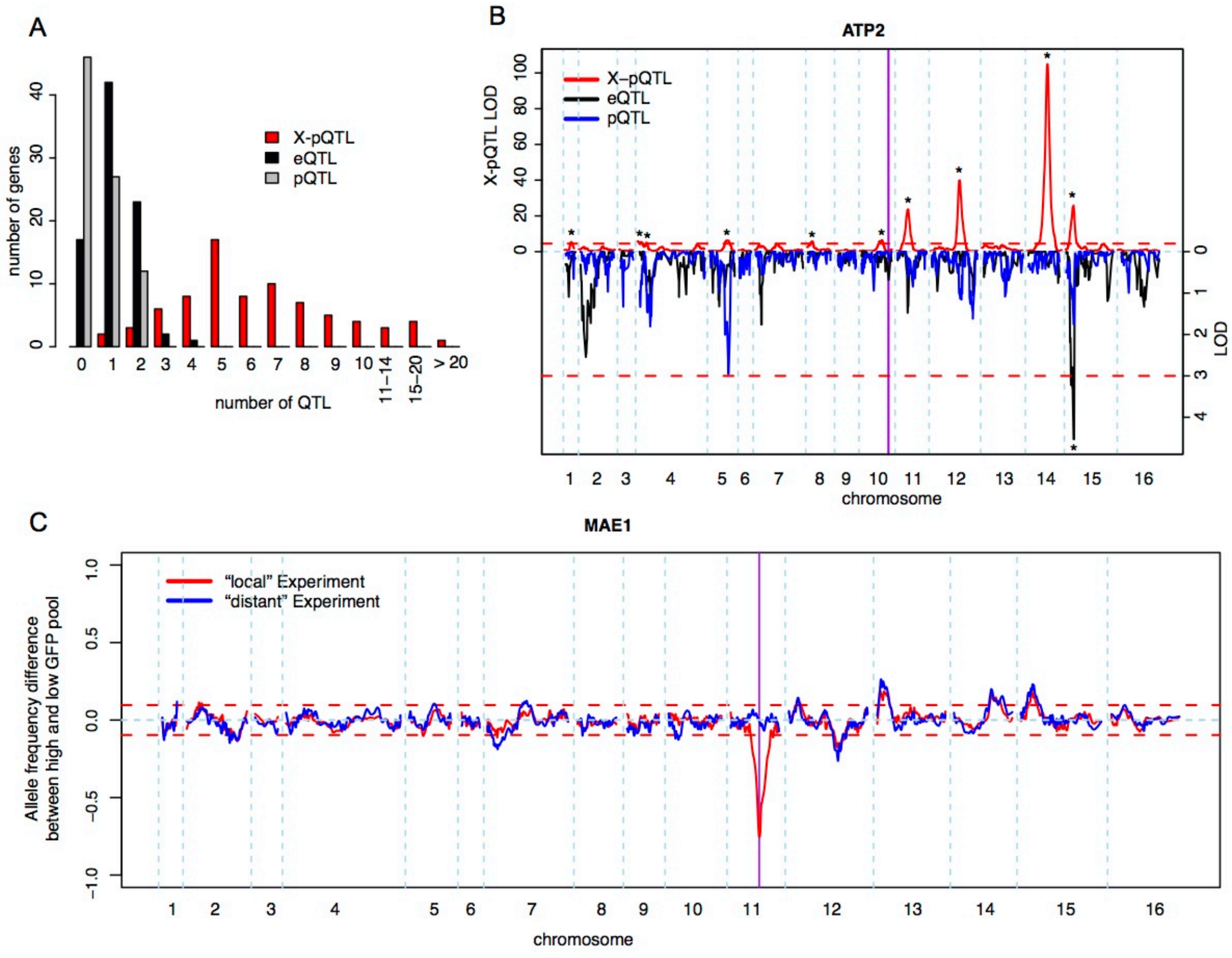

Figure 1 – Distant and local variation affects protein levels

A. Histogram showing the number of loci per gene among 85 genes with X-pQTL, eQTL and pQTL data. B. An example of protein and mRNA expression QTL for one representative gene (*ATP2*). Shown are X-pQTL LOD scores (top half) and eQTL / pQTL LOD scores (bottom half, inverted scale). The purple vertical line denotes the gene position. Red dashed horizontal lines indicate the genome wide significance thresholds. C. An example for a local X-pQTL in the gene *MAE1*. Shown is the difference in the frequency of the BY allele between the high and the low GFP population along the genome. Red dashed horizontal lines indicate the 99.99% quantile from the empirical "null" sort experiments. They are shown for illustration only and were not used for peak calling.

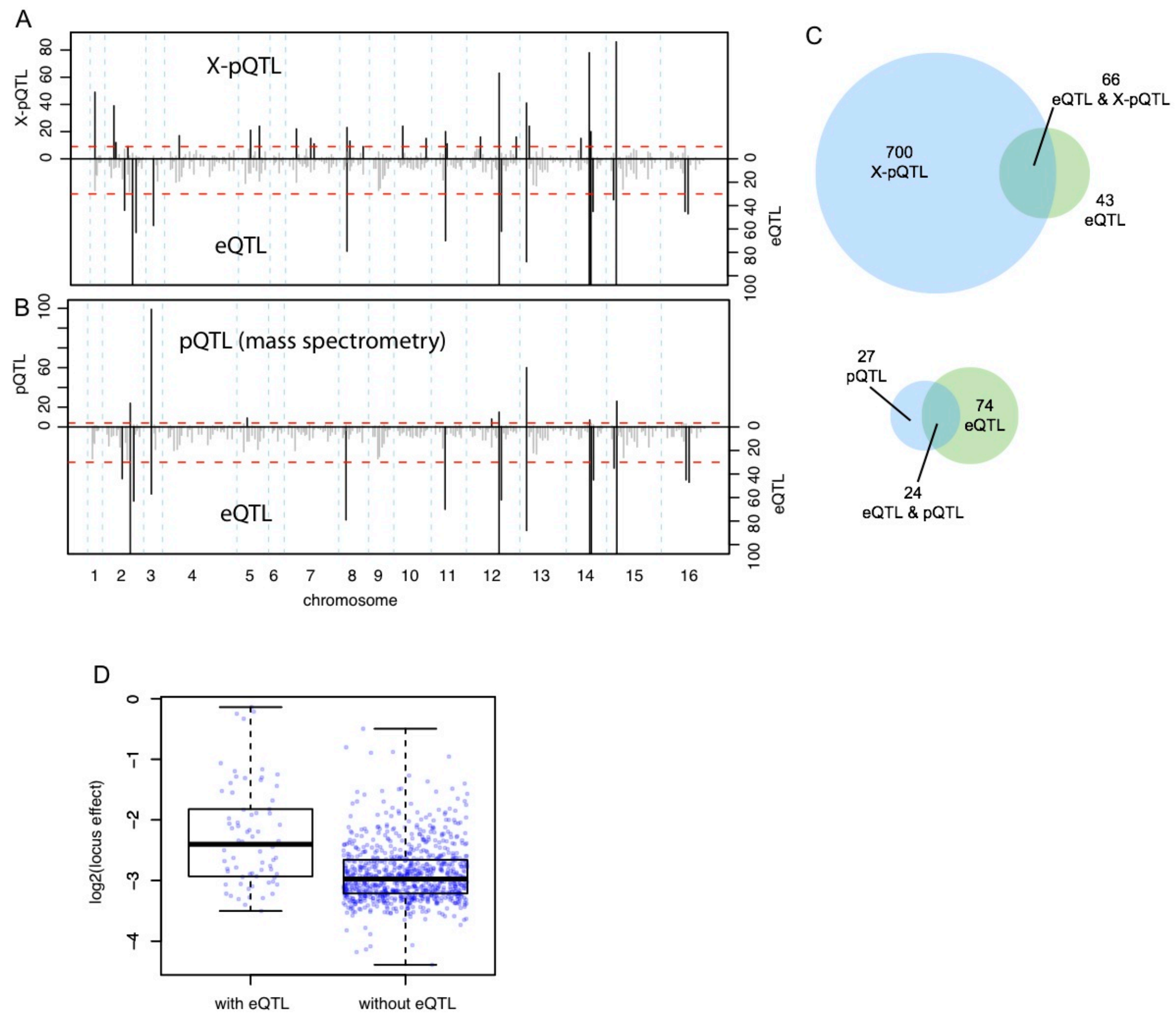

Figure 2 – X-pQTL hotspots and overlap with loci affecting mRNA abundance

A & B. Number of QTL along the genome. The genome was divided into 20 cM bins and in each bin the number of QTL was counted. Top half: X-pQTL, bottom half: eQTL. The red dashed lines correspond to the 95% quantiles of 100 datasets where QTL were distributed randomly across the genome. Bins where the QTL count exceeds this threshold are shown in black, others in grey. Note that the eQTL axes are truncated to permit easier visual comparison to X-pQTL data. A. X-pQTL (top) vs. eQTL (bottom). B. Mass-spectrometry based pQTL (top) vs. eQTL (bottom). The eQTL hotspot glu1 in Table 2 narrowly failed the permutation threshold in our re-analysis. The eQTL hotspots on chromosomes II and III (glu3, glu4, glu5) correspond to polymorphisms that do not segregate in our strains (in *AMN1*, *LEU2*, and *MAT*, respectively). The eQTL hotspot glu13 on chromosome XVI narrowly failed to reach significance in our data set. C. Overlap between eQTL and X-pQTL and between eQTL and pQTL. D. Distributions of X-pQTL effect sizes for X-pQTL with and without a corresponding eQTL. Effect sizes are shown as the allele frequency differences between the high and low GFP population.

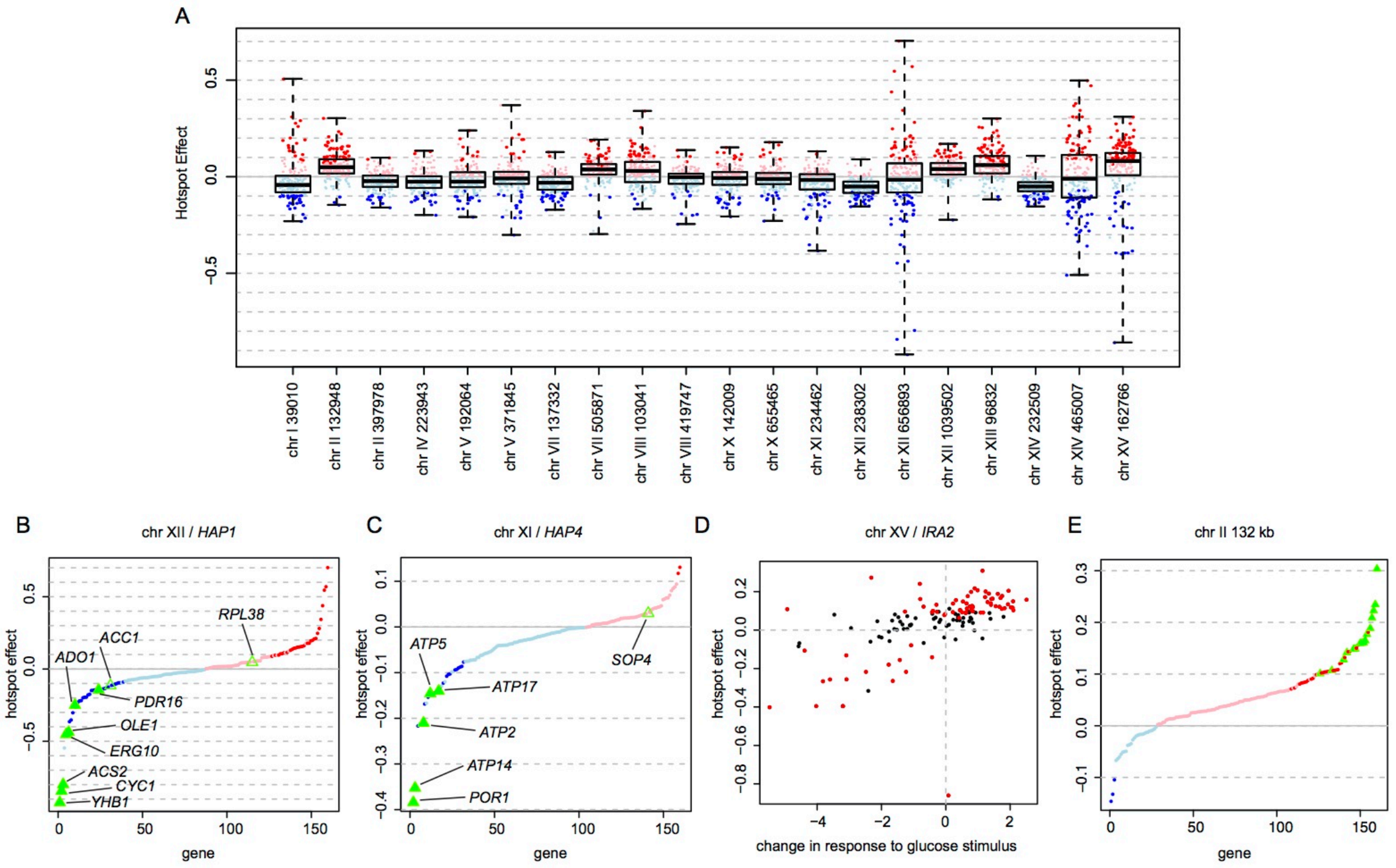

Figure 3 – Hotspot effects

A. The distribution of effects on all 160 individual genes for the 20 hotspots. Red (blue) indicates higher (lower) expression associated with the BY allele. Darker dots indicate significant X-pQTL. Superimposed boxplots show the median (central thick line), 25% and 75% percentile (boxes), and data extremes (whiskers). B & C. Effects of the *HAP1* and *HAP4* hotspots, with genes sorted by effect size. Green triangles indicate direct transcriptional targets of *HAP1* or *HAP4*. Filled triangles indicate significant X-pQTL. See text for details. D. Correlation of hotspot effects with expression changes triggered by glucose response. Red circles denote genes significantly regulated by the hotspot. E. Effects of chromosome II hotspot at position 132,948. Green triangles indicate genes with ribosomal and translation-related functions (See Supplementary Table S2 for gene information).

## Supplementary figures

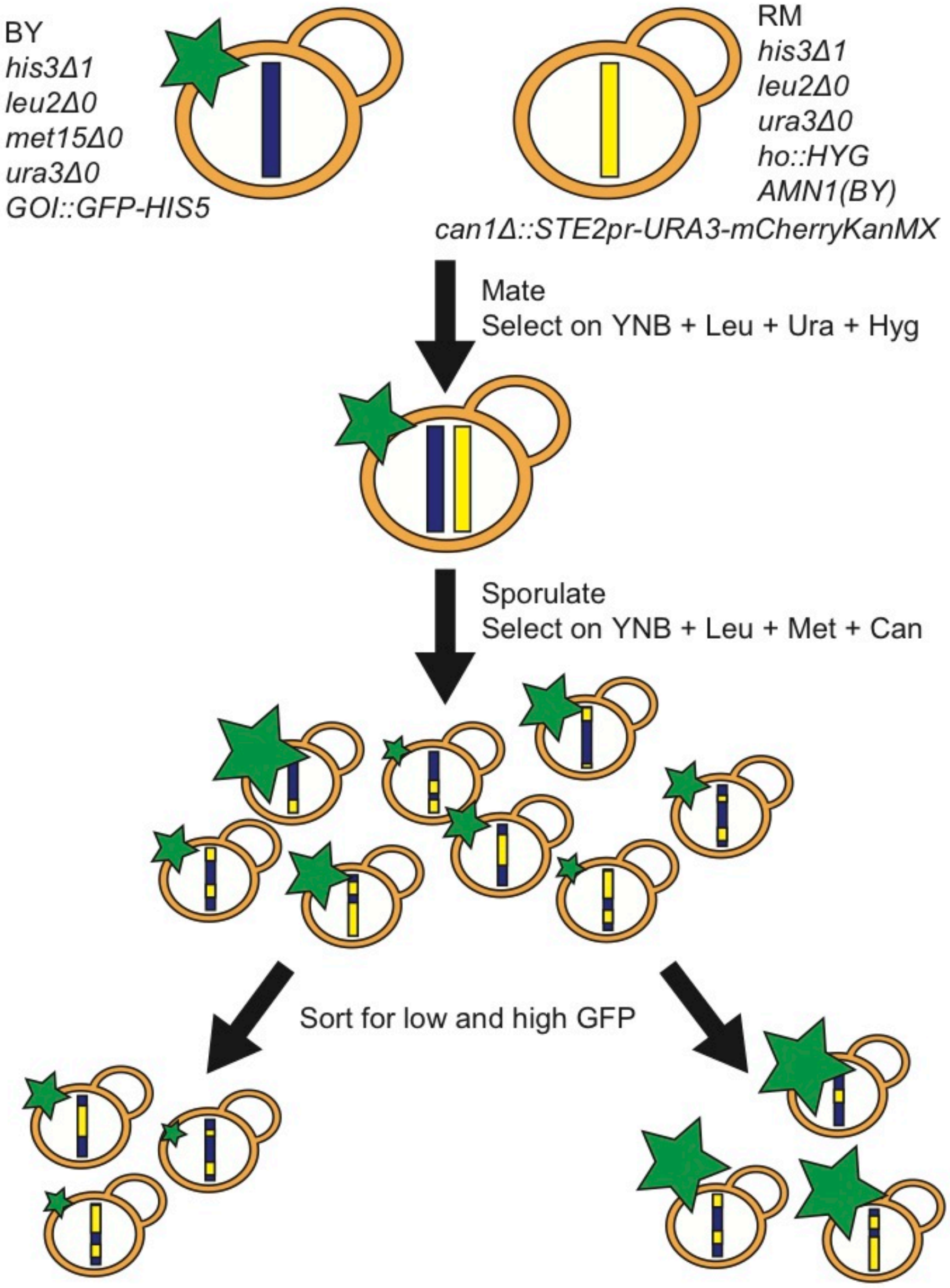

Supplementary

Supplementary Figure S1 – Overview of the experimental design

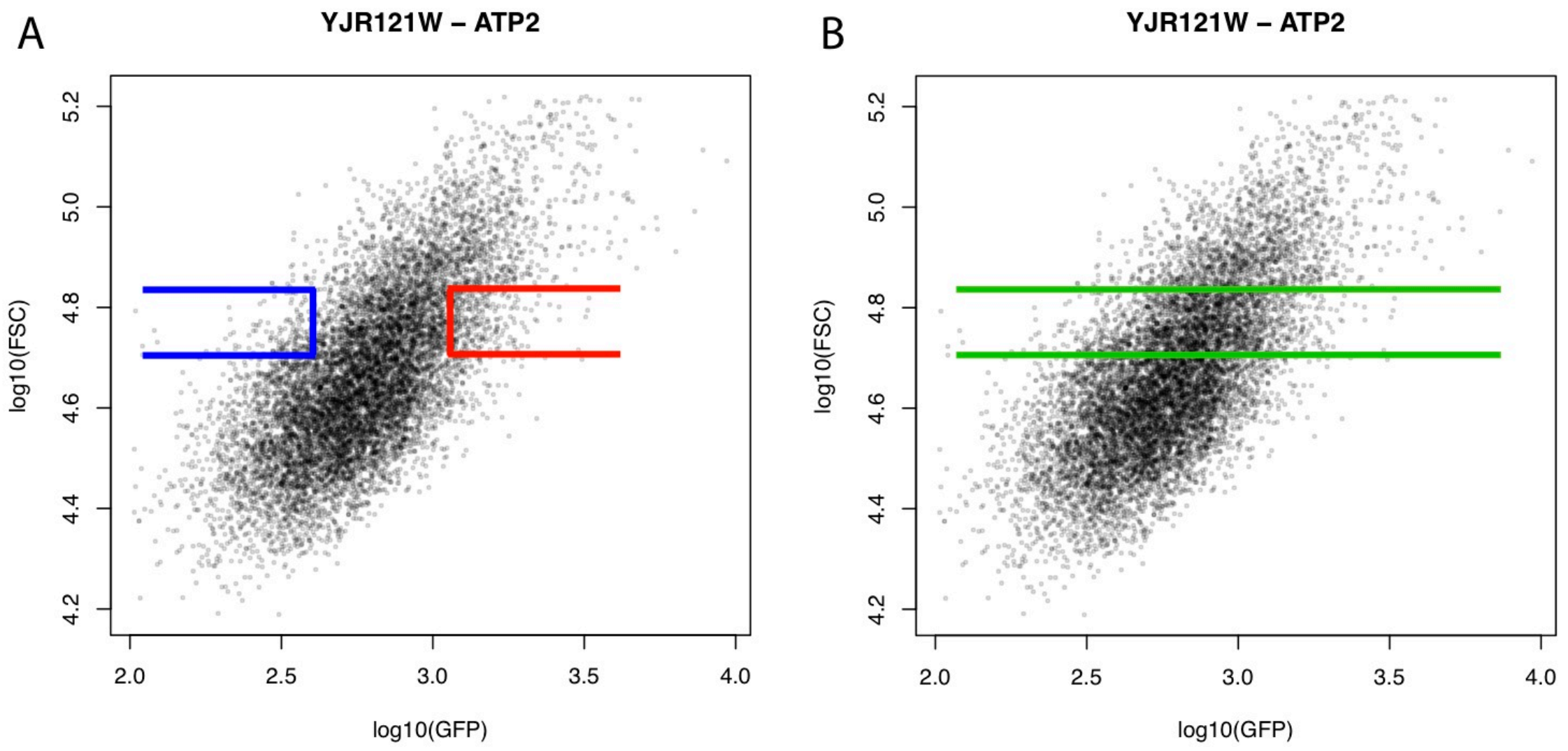

Supplementary Figure S2 – Illustration of FACS design

Shown is GFP intensity and forward scatter (FSC, a measure of cell size) recorded during FACS. The correlation between cell size and GFP intensity is clearly visible. The superimposed collection gates are an illustration, and do not show the actual gates used for this gene. A. The low GFP (blue) and high GFP (red) gates sample extreme levels of GFP within a defined range of cell sizes. B. For the "null" experiments, the same cell size range is collected, but without selecting on GFP.

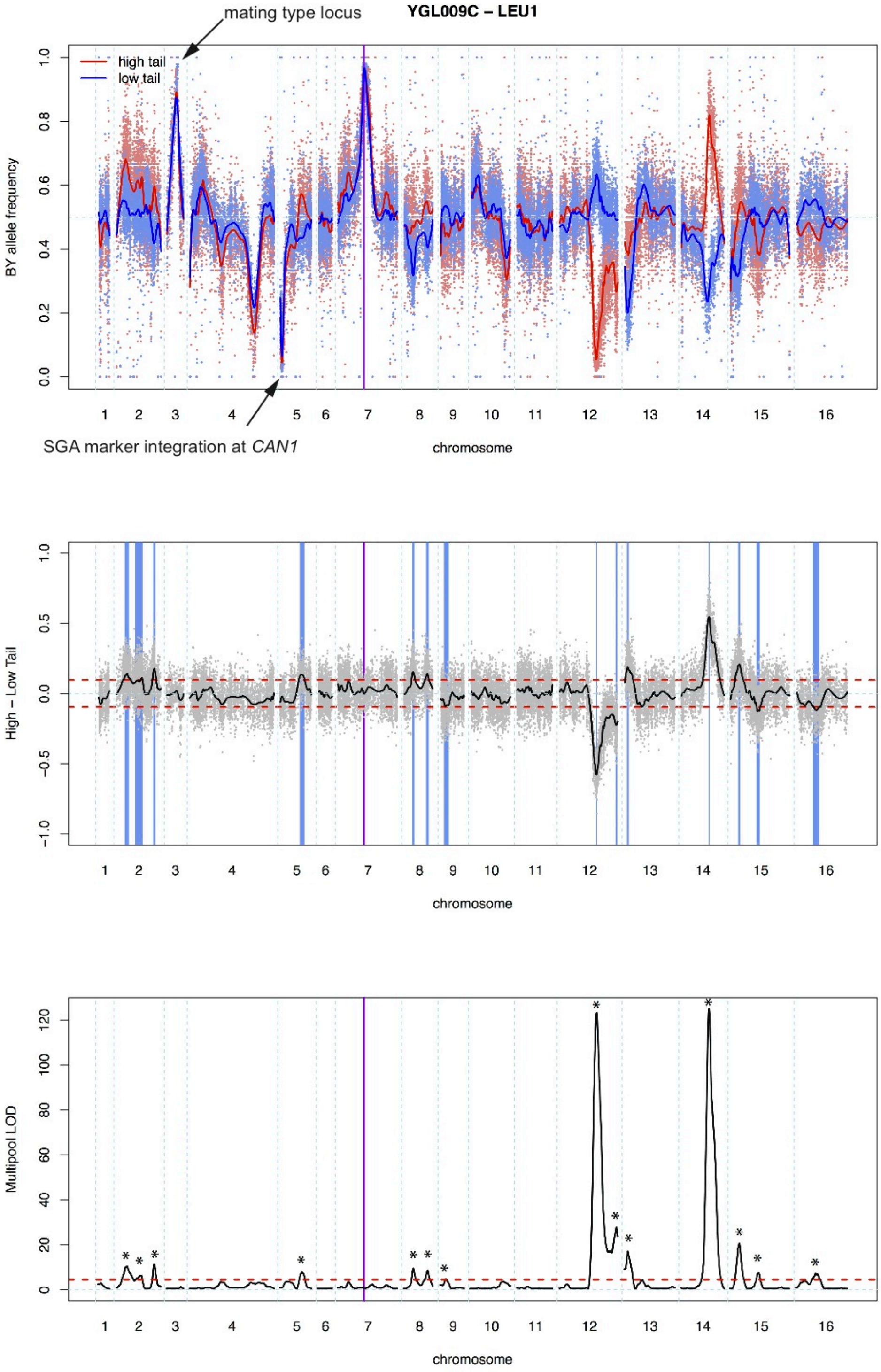

Supplementary Figure S3 – Sequence analyses and X-pQTL detection example

In all panels, physical genomic coordinates are shown on the x-axes. The position of the gene (*LEU1*) is indicated by the purple horizontal line.

Top panel: Frequency of the BY allele in the high (red) and low (blue) GFP population. SNPs are indicated by dots, and loess-smoothed averages as solid lines. Note the fixation for the BY allele in all segregants at the gene position as well as at the mating type locus on chromosome III, as well as the fixation for the RM allele at the SGA marker integrated at the *CAN1* locus on the left arm of chromosome V.

Middle panel: Subtraction of allele frequencies in the low from those in the high GFP population. SNPs are indicated by grey dots, with the loess-smoothed average indicated in black. Note that on average, there is no difference between the high and the low populations. Positive difference values correspond to a higher frequency of the BY allele in the high GFP population, which we interpret as higher expression being caused by the BY allele at that locus. The red horizontal lines indicate the 99.99% quantile from the empirical "null" sort experiments. They are shown for illustration only and were not used for peak calling. The blue vertical boxes indicate positions of genome-wide X-pQTL, with the width representing the 2-LOD drop interval.

Bottom panel: LOD scores obtained from MULTIPOOL. The red horizontal line is the genome-wide significance threshold (LOD = 4.5). Stars indicate X-pQTL called by our algorithm; these positions correspond to the blue bars in the middle panel. For this gene, 14 X-pQTL are called.

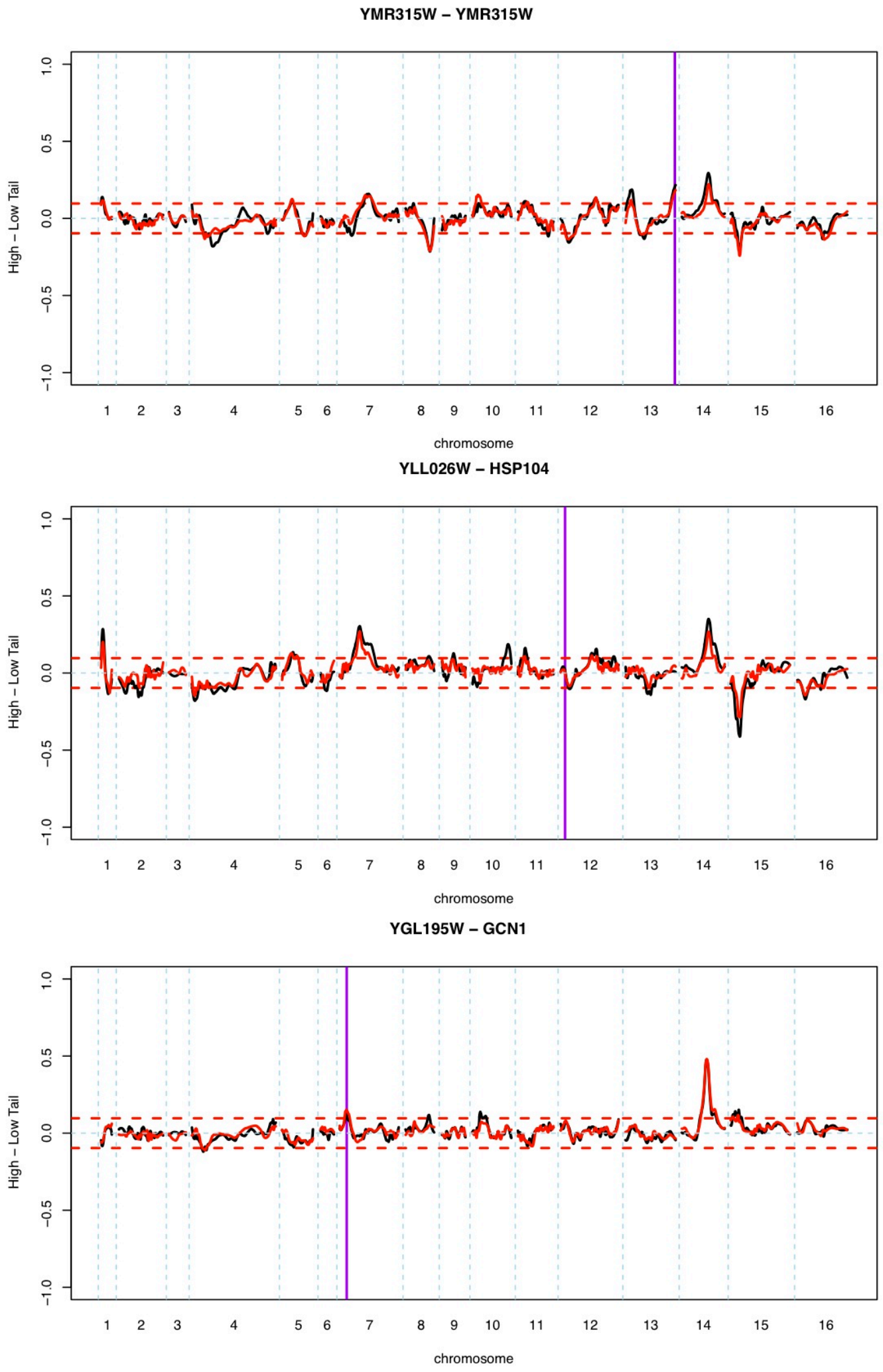

Supplementary Figure S4 – Reproducibility examples

Shown are allele frequency differences between the high and low GFP populations along the genome for three examples of replicates for three genes. The gene positions are indicated by purple vertical lines; note that *YMR315W* and *GCN1* were "local" experiments where peaks at the gene position are visible. The red horizontal lines indicate the 99.99% quantile from the empirical "null" sort experiments. Note the near-perfect agreement for strong X-pQTL, with some differences discernable at weaker loci.

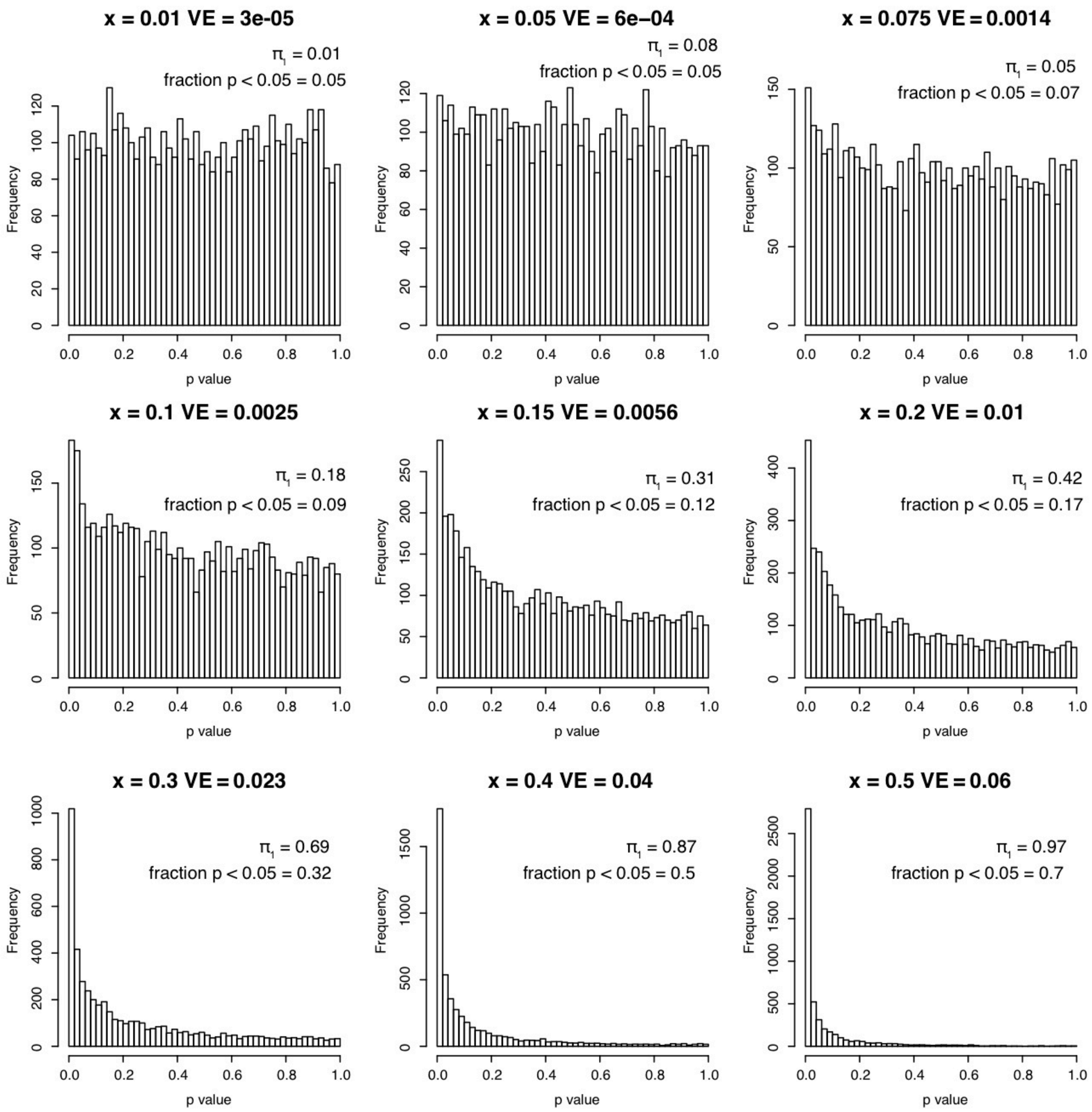

Supplementary Figure S5 – The impact of small effect sizes on the $\pi_1$ estimate

Each panel shows the p-value distribution obtained from 5,000 tests of a given effect size x, if two groups of 50 individuals each are compared using a T-test. The effect size x is given along with the corresponding variance explained (VE), the $\pi_1$ estimate, and the fraction of tests that achieved nominal significance ($p < 0.05$). Note that $\pi_1$ reaches 0.3 at VE = 0.5% – 1% (middle row, right columns). See Supplementary Note S2 for details.

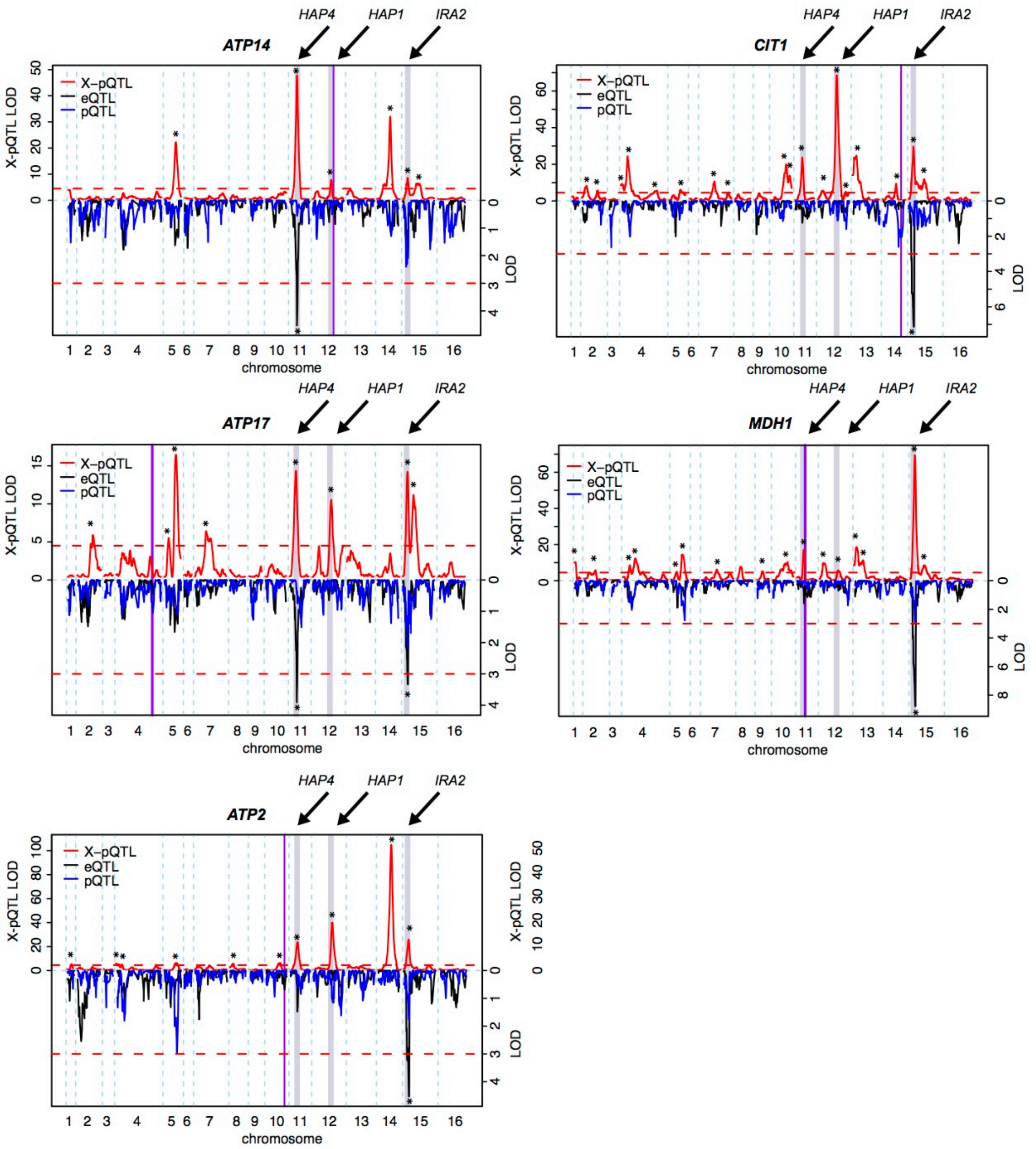

Supplementary Figure S6 – Genes regulated by the hotspots on chromosomes XI, XII, and XV involved in aerobic respiration

For each gene, we show the X-pQTL LOD scores along the genome in the top half of the plot, and the eQTL and pQTL LOD scores in the bottom half on an inverted scale. The hotspot locations are shown as grey bars labeled with the names of the causative genes. Purple vertical lines indicate the gene positions. Red dashed horizontal lines are significance thresholds. Stars indicate significant QTL.

# Supplementary Tables

Supplementary Table S1 – Information on genes used in the study; available as an EXCEL file

Supplementary Table S2 – Proteins affected by the hotspot on chromosome II

| Gene | Hotspot effect | Function | translation / ribosome related? |
|---|---|---|---|
| RPS17A | 0.30 | ribosomal protein small subunit | yes |
| TIF1 | 0.23 | translation initiation factor eIF4A | yes |
| PRT1 | 0.22 | translation initiation factor subunit eIF3b | yes |
| TEF4 | 0.21 | translational elongation factor | yes |
| RPS25A | 0.19 | ribosomal protein small subunit | yes |
| ADO1 | 0.18 | adenosine kinase | |
| UTP4 | 0.18 | Subunit of U3-containing 90S preribosome; involved in 18S rRNA production | yes |
| RPL9A | 0.16 | ribosomal protein large subunit | yes |
| RPL19A | 0.16 | ribosomal protein large subunit | yes |
| SUP45 | 0.16 | Polypeptide release factor eRF1 | yes |
| RPL21B | 0.16 | ribosomal protein large subunit | yes |
| LEU4 | 0.16 | leucine biosynthesis | |
| ILS1 | 0.15 | Isoleucine tRNA synthetase | |
| RPL13B | 0.15 | ribosomal protein large subunit | yes |
| CAM1 | 0.15 | transcription factor involved in ribosome biogenesis | yes |
| YOP1 | 0.15 | membrane traffic | |
| SUR4 | 0.14 | elongase involved in fatty acid and sphingolipid biosynthesis | |
| URA5 | 0.14 | pyrimidine biosynthesis | |
| RPL10 | 0.14 | ribosomal protein large subunit | yes |
| ADE3 | 0.13 | biosynthesis of purines, thymidylate, methionine, and histidine | |
| NOP58 | 0.13 | pre-rRNA processing, 18S rRNA synthesis, and snoRNA synthesis | yes |
| ILV6 | 0.12 | branched-chain amino acid biosynthesis | |
| SMI1 | 0.12 | cell wall synthesis | |
| PUB1 | 0.11 | Poly (A)+ RNA-binding protein required for mRNA stability | |
| TPI1 | 0.11 | Triose phosphate isomerase involved in glycolysis | |
| LHP1 | 0.11 | tRNA processing | |
| YIP3 | 0.11 | ER to Golgi transport | |
| TIF3 | 0.11 | Translation initiation factor eIF-4B | yes |
| TRX2 | 0.11 | cell redox homeostasis | |
| URA2 | 0.11 | pyrimidine biosynthesis | |
| ERG10 | 0.11 | ergosterol biosynthesis | |
| TRX1 | 0.10 | cell redox homeostasis | |
| CPR1 | 0.10 | cellular protein metabolism | |
| YLR413W | 0.10 | unknown | |
| DBP3 | 0.10 | rRNA processing | yes |
| YLR179C | 0.10 | unknown | |
| PFY1 | 0.09 | cytoskeleton organization | |
| LIA1 | 0.09 | cytoskeleton organization | |
| COX17 | 0.09 | copper transport, mitochondrial respiration | |
| PRS3 | 0.09 | nucleotide, histidine, and tryptophan biosynthesis | |
| TDH3 | 0.08 | glycolysis | |
| PDB1 | 0.08 | pyruvate dehydrogenase | |
| TPO1 | 0.08 | transmembrane transporter | |

| | | |
|---|---|---|
| PHO86 | 0.08 | ER to Golgi transport |
| ZWF1 | 0.07 | pentose phosphate pathway |
| WBP1 | 0.07 | protein glycosylation |
| CIT1 | -0.10 | citrate synthase, TCA cycle |
| HSP104 | -0.13 | chaperone |
| GPH1 | -0.15 | glycogen phosphorylase |

Supplementary Dataset S1 – List of X-pQTL; available as an EXCEL file

## Supplementary Note 1 – Reproducibility

We performed replication experiments for 27 genes. Of these, 22 were experimental replicates where the same segregant pool was thawed, grown and subjected to X-pQTL mapping at two different times. The remaining five replicates were done from independently generated crosses of the same gene. All replicates were processed several months apart from each other. We randomly assigned one of the two replicates as the detection set, and asked to what extent the X-pQTL identified in the detection set recurred in the validation set.

Across the 27 detection experiments, we discovered 240 X-pQTL at genome-wide significance. We first asked whether the validation experiment had an allelic effect in the same direction (i.e. with higher expression associated with either the BY or the RM allele), irrespective of significance in the validation set. The direction of effect was concordant at 234 (97.5%) of the X-pQTL. We next asked what fraction of X-pQTL was reproduced at genome-wide significance in the replication set. We found that peaks with higher LOD scores were more likely to be reproduced, ranging from 58% replication at LOD ≥ 4.5 to perfect replication at higher LODs (Supplementary Note Table 1). Genome-wide significance in both datasets is a strict criterion for the small effects detected in our study. We therefore also employed a relaxed replication criterion, where we required the validation set to show an allele frequency difference of at least 0.05 (~ half that required for genome-wide significance) and a concordant direction of allelic effect. Using this criterion, 80% of loci reproduced at detection LOD ≥ 4.5, rapidly approaching perfect replication at higher LOD thresholds (Supplementary Note Table 1).

In sum, we found that loci with strong effect virtually always reproduce. Loci of smaller effect sometimes fail to reach genome-wide significance in a replication experiment, likely due to stochastic variation of the influence of small effect loci, perhaps due to minute differences in the experimental conditions (e.g. selection strength, or small variations in temperature or cell density). Notably, even loci of small effect are still concordant in their direction of effect in the vast majority of cases. Supplementary Figure S4 illustrates these patterns for three genes.

Supplementary Note Table 1 – Reproducibility statistics

| LOD threshold in detection set | X-pQTL | Replicated at genome-wide significance | Replicated at reduced significance & concordant direction* |
|---|---|---|---|
| 4.5 | 240 | 139 (58%) | 191 (80%) |
| 5 | 216 | 128 (59%) | 176 (81%) |
| 10 | 86 | 69 (80%) | 78 (91%) |
| 20 | 30 | 29 (97%) | 30 (100%) |
| 50 | 10 | 10 (100%) | 10 (100%) |

*see text for details

## Supplementary Note 2 – Influence of small effect sizes on eQTL detection

We used the statistic $\pi_1$ [1], to estimate the fraction of X-pQTL that have an underlying eQTL, and obtained (after correcting for random overlap, see Methods), an estimate of ~32%. Taken at face value, the estimate suggests that 68% of X-pQTL are due to genetic variation that specifically influences posttranscriptional regulation, without affecting mRNA levels. We sought to explore the alternative explanation that $\pi_1$ might underestimate the fraction of true positive tests if effect sizes are small. We note that $\pi_1$ is designed to be a lower bound of the fraction of true positive tests in a multiple testing scenario [1], but we here sought to quantify this effect in more detail.

We performed simulations of the situation where a single position in a genome is tested for a difference in phenotypes (such as mRNA levels for a given gene) between haploid individuals of either of two genotypes. The test is a T-test of the phenotypes in the two groups. To form the two groups, we randomly sampled 50 phenotypes each from a normal distribution with standard deviation = 1. Individuals in the first group had mean phenotype = 0, while those in the second group had a higher mean phenotype = x. For $x \ll 1$, the expected variance explained by the group difference (i.e., the "eQTL" effect size) is $x^2/4$. For each x, we generated 5,000 sets of two groups (resembling 5,000 "genes"), performed a two-sided T-test in each set, recorded the p-values and calculated $\pi_1$ from the distribution of the 5,000 p-values. Importantly, these simulations probe the behavior of $\pi_1$ in the situation where *every* test is truly positive because x is never equal to 0. If power were sufficiently high, $\pi_1$ should equal 1 irrespective of x. Any lower value of $\pi_1$ is a consequence of low power due the relatively small sample size of 100 individuals.

Supplementary Figure S5 shows that for high x, $\pi_1$ indeed approaches 1. At smaller x however, $\pi_1$ is reduced along with the power to detect individual "genes" to be significant. For very small x, power is at chance level, and $\pi_1$ estimates are near zero. Importantly, a $\pi_1$ of 0.3 – 0.4 (as seen in our actual data) is reached at values of x that correspond to an effect size of 0.5 – 1% of variance explained (center row). Thus, rather than requiring wide-spread posttranscriptional consequences of genetic variation, the observed estimate of $\pi_1$ can also be explained by many (perhaps most or even all) X-pQTL having true, but sufficiently small effects on mRNA levels.

1. Storey, J. D. & Tibshirani, R. Statistical significance for genomewide studies. *Proceedings of the National Academy of Sciences* **100,** 9440–9445 (2003).